\documentclass[12pt]{article}

\newcommand{\be}{\begin{equation}}
\newcommand{\ee}{\end{equation}}
\newcommand{\bea}{\begin{eqnarray}}
\newcommand{\eea}{\end{eqnarray}}
\newcommand{\ba}{\begin{array}}
\newcommand{\ea}{\end{array}}

\newcommand{\al}{\alpha}

\newcommand{\de}{\delta}
\newcommand{\ep}{\epsilon}
\newcommand{\si}{\sigma}
\newcommand{\la}{\lambda}

\newcommand{\ze}{\zeta}

\newcommand{\tha}{\theta}
\newcommand{\Ups}{\Upsilon}

\newcommand{\ttha}{\tilde{\tha}}
\newcommand{\tQ}{\tilde{Q}}
\newcommand{\pa}{\partial}
\newcommand{\rar}{\rightarrow}
\newcommand{\non}{\nonumber}
\newcommand{\cN}{\mathcal{N}}
\newcommand{\half}{\mbox{$\frac{1}{2}$}}
\newcommand{\fourth}{\mbox{$\frac{1}{4}$}}

\begin{document}

\vspace*{-6mm}

\rightline{\vbox{\footnotesize
\hbox{DAMTP-1999-149}
\hbox{\tt hep-th/9910160}
}}

\vspace{3mm}

\begin{center}
{\Large\sf (Super)conformal many-body quantum mechanics \\ with extended supersymmetry }

\vskip 5mm

Niclas Wyllard\footnote{\tt N.Wyllard@damtp.cam.ac.uk} \vspace{5mm}\\
{\em DAMTP, University of Cambridge,\\ Silver Street, Cambridge CB3 9EW, UK}

\end{center}
 
\vskip 5mm
 
\begin{abstract}
\noindent We study $\cN=4$ supersymmetric quantum-mechanical
many-body systems with $M$ bosonic and $4M$ fermionic degrees of
freedom. We also investigate the further
restrictions of conformal and superconformal invariance. 
 In particular, we construct conformal $\cN=4$ extensions of the $A_{M-1}$
Calogero models, which  for
generic values of the coupling constant are not $\mathrm{SU}(1,1|2)$
superconformal. This class of models is also extended to arbitrary
(even) $\cN$.  We give both hamiltonian and (classical) lagrangean
formulations. In the latter case we use both  component and $\cN=4$ superfield  formulations. 
\end{abstract}

\setcounter{equation}{0}

\section{Introduction}
Recently there has been increased interest in supersymmetric quantum-mechanical models. Contrary to the situation in higher dimensions, such models have been much less studied. One recent application is to black hole physics
\cite{Gibbons:1998,Blum:1999,Michelson:1999b,Gutowski:1999}. A related
issue is the still incompletely understood ${\rm adS}_2/{\rm CFT}_1$
correspondence \cite{Maldacena:1997}. In the case of
black holes, most work has so far been concerned with $\cN=4$ models with $4M$
bosonic and $4M$ fermionic coordinates, and  general results for such
models have been obtained
\cite{Coles:1990,Michelson:1999a,Hull:1999}. The emphasis has been on
which (sigma-model) metrics are consistent with supersymmetry and the
properties of the resulting geometries. Our focus is slightly
different; we discuss models with $M$ bosonic and $4M$ fermionic
degrees of freedom, take the metric to be flat and study the constraints on the potential coming from
supersymmetry. We also investigate the constraints arising from adding
more symmetry such as translational invariance,
conformal invariance and superconformal invariance.  We do
not have any particular application in mind, although $\cN=4$
supersymmetric superconformal Calogero models have been conjectured \cite{Gibbons:1998} to provide a
microscopic description of four-dimensional extremal
Reissner-Nordstr\"om  black holes. The Calogero models
\cite{Calogero:1969a} and their generalisations comprise a particular class of many-body quantum-mechanical models that have been intensely studied
over the years.  These models have appeared in various areas of
theoretical physics, ranging from problems in condensed matter physics to
Seiberg-Witten theory. For reviews with extensive
lists of references to the early literature, see
\cite{Olshanetskii:1981} (for reviews on the connection to
Seiberg-Witten theory, see e.g.~\cite{Mironov:1999}). It is well known that the
Calogero systems are intimately connected with the semi-simple Lie
algebras. For every (semi-simple) Lie algebra there is an associated
Calogero system. It is perhaps less widely known that the conditions
can be weakened. Recently it has been shown \cite{Haschke:1998b,Bordner:1999a} that one actually does not need a root system associated to a Lie algebra to construct a Calogero model. It is sufficient to have a root system associated to any finite reflection group (Coxeter group); only when the root system is crystallographic can one  associate it to a Lie algebra and the Coxeter group is then called the Weyl group.
The Calogero systems are integrable (see e.g.~\cite{D'Hoker:1998,Bordner:1999a} and references therein) and, for the
cases with discrete spectrum, exactly
solvable. By exactly solvable we mean the condition that it should be
possible to obtain the eigenfunctions in an ``algebraic'' way. This
has been shown  using  various different approaches, see
e.g.~\cite{Brink:1992,Ruehl:1995,Gurappa:1999}. An interesting feature
of the $A_{M-1}$ Calogero models is that they are translational- and conformal-invariant. The two-particle case coincides (after removing the centre-of-mass motion) with the model of conformal mechanics studied in \cite{deAlfaro:1976}.   
Supersymmetric extensions of the Calogero models with $\cN=2$ supersymmetry have also been constructed \cite{Freedman:1990,Brink:1993,Brink:1997,Bordner:1999b}. So far the supersymmetric models have not had as many applications as the bosonic models. 
 The models constructed in \cite{Freedman:1990} are also
superconformal; the superconformal algebra being
$\mathrm{osp}(2|2)\cong \mathrm{su}(1,1|1)$. The relative motion of
the two-particle case was studied before in
\cite{Akulov:1984}. In \cite{Ivanov:1989b,Pashnev:1986,Akulov:1999} (see also
\cite{deAzcarraga:1998}) an $\cN=4$ superconformal extension of the
conformal quantum mechanics model was constructed (a related
development is \cite{Zhou:1999}). The superconformal
group in this case is $\mathrm{SU}(1,1|2)$. This result has not been
extended to the many-body case.

In the next section we investigate (using the quantum hamiltonian
formalism \cite{Witten:1981}) the restrictions of $\cN=4$ supersymmetry, conformal
invariance and $\mathrm{SU}(1,1|2)$ superconformal symmetry. We 
first discuss the one-particle case and then move on to the
many-body case and derive general results. We  concentrate on the
$A_{M-1}$ Calogero models, but our results are applicable also to other
cases. We show that it is possible to construct conformal $\cN=4$
extensions of the $A_{M-1}$ Calogero models, which are $\mathrm{SU}(1,1|2)$
superconformal only for a particular value of the coupling
constant. Furthermore, we  show that (given certain assumptions) for $M\!>\!2$ and generic values
of the coupling constant there are no natural $\mathrm{SU}(1,1|2)$
superconformal extensions of the $A_{M-1}$ Calogero models. In section~\ref{slag} we  present a similar discussion employing the language of the
classical lagrangean formalism \cite{Salomonson:1982}. We use both superfield and component
formulations. We  also briefly discuss the connection between the
classical lagrangean approach and the quantum hamiltonian one. We end
with a short discussion of the possible relevance of our results to black
hole physics and some open questions. 

\setcounter{equation}{0}
\section{Quantum hamiltonian formulation} \label{sham}
We assume the $\cN=4$ supersymmetry algebra to be of the form
\be
{}[Q_a,Q^{\dagger b}]_+ = 2\de_{a}^b H\,, \quad [Q_{a},Q_{b}]_+ = 0\,,
\quad [Q^{\dagger a},Q^{\dagger b}]_+ = 0\,, \label{susyalg}
\ee
where $a,b=1,2$. In other words, we use a complex formalism.  
Some of our conclusions may be altered if the supersymmetry algebra is
changed, i.e. if  central charges are allowed or if more general
supersymmetry algebras are considered (such as the ones in
\cite{Gibbons:1993,Hull:1999}). In this section we will
investigate the restrictions on the potential resulting from requiring an $\cN=4$ symmetry in the form of the supersymmetry algebra (\ref{susyalg}). We will also discuss the restrictions coming from demanding conformal and superconformal invariance.

\subsection{Preliminaries: one-body models}
The supercharges are $Q_a$ and their hermitean conjugates. We sometimes use the notation $Q\equiv Q_1$ and $\tilde{Q}\equiv Q_2$ to reduce the number of indices. The discussion below of the one-body case is in part a review (of \cite{Pashnev:1986,Akulov:1999}), but it is presented in such a way as to facilitate the extension to the many-particle case to be discussed later. 
We will denote the bosonic coordinate by $x$ and use the concrete realisation $\tha$, $\ttha$, $\pa_{\tha}$ and $\pa_{\ttha}$ for the fermionic coordinates. On general grounds the supercharges can be taken to be of the form:
\be
Q=\tha\left(p -iW^{[0]}(x) - i W^{[1]}(x)\ttha\frac{\pa}{\pa\ttha}\right) \,, \quad \tQ=\ttha\left(p -iW^{[0]}(x) - i W^{[1]}(x)\tha\frac{\pa}{\pa\tha}\right)\,, \label{1scharges}
\ee
together with their hermitean conjugates (using $\tha^{\dagger} = \frac{\pa}{\pa \tha}$ and $(\xi\ze)^{\dagger} = \ze^{\dagger}\xi^{\dagger}$).
The supersymmetry algebra (\ref{susyalg}) is satisfied if the following equation is  satisfied ($\pa \equiv \frac{d}{dx}$)
\be
2\pa W^{[0]} -2W^{[0]} W^{[1]}+ \pa W^{[1]} - W^{[1]} W^{[1]}=0\,. \label{1constr}
\ee
The hamiltonian then becomes
\be
H= \half p^2 + \half(W^{[0]})^2 - \half\pa W^{[0]} +  \pa W^{[0]}
[\tha\frac{\pa}{\pa\tha} + \ttha\frac{\pa}{\pa\ttha}] +  \pa
W^{[1]}\tha\frac{\pa}{\pa\tha}\ttha\frac{\pa}{\pa\ttha}\,. \label{1ham}
\ee
We will now restrict ourselves to conformal models. Such models
satisfy $[D,H]= -iH$, where $D=-\fourth [x,p]_+$
\cite{deAlfaro:1976}. To regain the $\mathrm{OSp}(2|2)$ superconformal
mechanics of  \cite{Akulov:1984} when restricting to the
$\cN=2$ sub-sector, we have to set $W^{[0]} = \frac{\nu}{x}$. Somewhat
surprisingly, for this choice of $W^{[0]}$ there are two solutions to the
constraint (\ref{1constr}) preserving conformal invariance, namely: $W^{[1]}=-\frac{1}{x}$ and $W^{[1]}=-\frac{2\nu}{x}$. Thus, there are two different conformal $\cN=4$ supersymmetrisations of conformal mechanics (or equivalently, of the relative motion of the $A_1$ Calogero model).
The corresponding hamiltonians are
\bea
H_{1} &=& \half p^2 + \frac{1}{2x^2}\left(\nu^2 + \nu  - 2\nu[\tha\frac{\pa}{\pa \tha} +  \ttha\frac{\pa}{\pa \ttha}] + 2\tha\frac{\pa}{\pa \tha} \ttha\frac{\pa}{\pa \ttha}\right)\,, \non \\
H_{2} &=& \half p^2 + \frac{1}{2x^2}\left(\nu^2 + \nu  - 2\nu[\tha\frac{\pa}{\pa \tha} +  \ttha\frac{\pa}{\pa \ttha}] + 4\nu\tha\frac{\pa}{\pa \tha} \ttha\frac{\pa}{\pa \ttha}\right) \non \\ &=& \half p^2 + \half \frac{\nu(\nu + [\tha,\pa_{\tha}][\ttha,\pa_{\ttha}])}{x^2} \,. \label{1hams}
\eea
Although both models in (\ref{1hams}) are conformal, only the first has
$\mathrm{su}(1,1|2)$ as its superconformal algebra for generic $\nu$. This can be seen by making a general Ansatz for the
generators of special supersymmetries $S$, $\tilde{S}$ and their
hermitean conjugates. The $\mathrm{su}(1,1|2)$ superconformal algebra (see
appendix) is satisfied if $S=\tha x$, $\tilde{S} = \ttha x$ and
$xW^{[1]}=-1$. The other generators of $\mathrm{su}(1,1|2)$ are then given by
\be
J_1 = -\half (\tha\pa_{\ttha} + \ttha\pa_{\tha})\,, \quad J_2 = -\mbox{$\frac{i}{2}$}(\tha\pa_{\ttha} - \ttha\pa_{\tha})\,, \quad J_3 = \half(\ttha\pa_{\ttha} - \tha\pa_{\tha})\,.
\ee
Furthermore, the central element $T$ is given by $T=xW^{[0]}-\half$. Thus, there is a unique  $\mathrm{SU}(1,1|2)$ conformally invariant
model. For this model $T=\nu -\half$. The second model above is
$\mathrm{SU}(1,1|2)$ superconformal only when $\nu=\half$, in which
case it coincides with a special case of the first model (with
$T=0$). For other values of $\nu$ the conformal and supersymmetry
generators belong to some other superconformal algebra. The free theory is not
$\mathrm{SU}(1,1|2)$ superconformal. The above $\mathrm{SU}(1,1|2)$ superconformal model
is, however, ``on-shell''-dual to a free $\cN=4$ theory with a complex
bosonic coordinate (2 real ones) \cite{Ivanov:1989b}. Let us also
mention that for the first model above there is a simple extension to
arbitrary (even) $\cN$ \cite{Akulov:1999}, i.e arbitrary number of
supersymmetries (there is no restriction on the number of
supersymmetries in one dimension since there is no notion of
spin). The arbitrary-$\cN$ models have the following supercharges \cite{Akulov:1999}
\be
Q_a = \tha_a(p - i \frac{\nu}{x} + i\frac{1}{x}\sum_{c\neq
a}\tha_c\pa_{\tha_c})\,, \quad Q^{\dagger a} = \pa_{\tha_a}(p + i
\frac{\nu}{x} - i\frac{1}{x}\sum_{c\neq a}\tha_c\pa_{\tha_c})\,,
\ee
where $a,c=1,\ldots,\frac{\cN}{2}$. 
These models are also superconformal; the superconformal algebra being
$\mathrm{su}(1,1|\frac{\cN}{2})$ (see appendix). The other generators of
$\mathrm{su}(1,1|\frac{\cN}{2})$ are given by: $S_a =
\tha_a x$, $S^{\dagger a}=\pa_{\tha_a}x$, and
\be
 J_{a}{}^{b} =
\tha_a\pa_{\tha_b}\;\;(a\neq b)\,, \quad J_{a}{}^{a} =
\tha_a\pa_{\tha_a} - \frac{2}{\cN}\sum_{c}\tha_c\pa_{\tha_c}\,,
\quad U=\half\sum_{c}\tha_c\pa_{\tha_c}\,.
\ee
For the second model in (\ref{1hams}) a similar extension to arbitrary (even) $\cN$ can be constructed by taking the supercharges to be of the form
\be
Q_a= \tha_a(p - iW^{[\frac{\cN-2}{2}]}(x)\prod_{c\neq a} [\tha_c,\frac{\pa}{\pa\tha_c}])\,, \quad Q^{\dagger a} = \pa_{\tha_a}(p + iW^{[\frac{\cN-2}{2}]}(x)\prod_{c\neq a} [\tha_c,\frac{\pa}{\pa\tha_c}]) \,,
\ee
where $a,c=1,\ldots \frac{\cN}{2}$ (we use a slightly different normalisation for $W^{[1]}$ than before).
The corresponding hamiltonian is obtained from $[Q_a,Q^{\dagger b}]_+= 2 \de_a^b H$ and becomes
\be
H=\half p^2 +  \half(W^{[\frac{\cN-2}{2}]})^2 + \half \pa W^{[\frac{\cN-2}{2}]} \prod_{c}[\tha_c,\frac{\pa}{\pa\tha_c}] \,.
\ee
In particular, for $W^{[\frac{\cN-2}{2}]}=-\frac{\nu}{x}$ we get
\be
H= \half p^2 +   \half \frac{\nu(\nu +
\prod_{c=1}^{\cN/2}[\tha_c,\frac{\pa}{\pa\tha_c}])}{x^2} \,.
\ee 
Notice that $(\prod_{c}[\tha_c,\frac{\pa}{\pa\tha_c}])^2 =1$. 

\subsection{Extension to many-body models}
We will now discuss the extension of the above results to the
many-body case. The coordinates are $x_i$, $\tha_i$, $\ttha_i$,
$\pa_{\tha_i}$ and $\pa_{\ttha_i}$, where $i=1,\ldots,M$. Here, and
throughout the paper, we will assume that the hamiltonians are
invariant under permutations of the coordinates. We take the
supercharge $Q$ to be of the form
\be
Q=\sum_j\tha_j\left(p_j -iW^{[0]}_{j}(x_k) - i\sum_{nm}
W^{[1]}_{jnm}(x_k)\ttha_n\frac{\pa}{\pa\ttha^m}\right)\,,  \label{scharges}
\ee
with a similar expression for $\tilde{Q}$. 
This is not the most general choice, but it is a natural extension of the supercharges used to construct $\cN=2$ models \cite{Freedman:1990,Shastry:1993,Brink:1997,Bordner:1999b}.
The supersymmetry algebra (\ref{susyalg}) is satisfied if the following conditions are fulfilled
\bea
&& W^{[0]}_i = \pa_i W^{[0]}\,, \qquad W^{[1]}_{ijk} = \pa_i\pa_j\pa_k W^{[1]} \,, \non \\
&& \sum_l W^{[1]}_{li[n}W^{[1]}_{m]jl} = 0  \,, \non \\
&& \pa_j\pa_k \hat{W}^{[0]} = \sum_l W^{[1]}_{ljk}\pa_l\hat{W}^{[0]}\,, \quad (\hat{W}^{[0]} := W^{[0]} + \half\sum_n \pa_n\pa_n W^{[1]} )\,, \label{constr}
\eea
and the hamiltonian is then given by 
\bea
H &=& \half\sum_i [p_i^2 + (\pa_i W^{[0]})^2 - \pa^2_{i}W^{[0]} ] + \sum_{i,j}(\tha_i\pa_{\tha_j} + \ttha_i\pa_{\ttha_j})\pa_{i}\pa_{j}W^{[0]} \non \\ && + \,\sum_{ijnm}\tha_i\pa_{\tha_j}\ttha_n\pa_{\ttha_m}\pa_{i}\pa_{j}\pa_{n}\pa_{m}W^{[1]} \,.
\eea
One solution to the last constraint in (\ref{constr}) is the trivial one: $\hat{W}^{[0]}=0$, i.e. $W^{[0]} =
-\half\sum_n \pa_n\pa_n W^{[1]}$. This provides a possible way to construct $\cN=4$ extensions of known $\cN=2$ models, e.g.~the Calogero models. We would like to stress that one also has to check that the other conditions in (\ref{constr}) hold. With $\hat{W}^{[0]}=0$, the supercharge $Q$  takes the form 
\be
Q=\sum_j\tha_j\left(p_j  - \frac{i}{2}\sum_{n,m} \pa_j\pa_n\pa_m W^{[1]} [\ttha_n,\frac{\pa}{\pa\ttha^m}]\right) \,,
\ee
with a similar expression for $\tQ$.
The $A_{M-1}$ Calogero models have $W^{[0]}= {\rm \frac{\nu}{2} }\sum_{i\neq j} \ln|x_i-x_j|$. If we set
\be
 W^{[1]} =  -\frac{\nu}{2}\sum_{i\neq j} \left[(x_i-x_j)^2\ln|x_i-x_j|
- \frac{3}{2}(x_i-x_j)^2\right]\,, \label{w1cal}
\ee
it can readily be checked that all conditions in (\ref{constr}) are fulfilled. The resulting hamiltonian becomes 
\be
H = \half\sum_i p_i^2 + \sum_{i < j}\frac{\nu(\nu + K_{ij})}{(x_i-x_j)^2}\,, \label{ccal}
\ee
where $K_{ij}
= \fourth[(\tha_i-\tha_j),(\pa_{\tha_i}-\pa_{\tha_j})][(\ttha_i-\ttha_j),(\pa_{\ttha_i}-\pa_{\ttha_j})]$. The operator $K_{ij}$ is an exchange operator satisfying: $\tha_i K_{ij} = K_{ij}\tha_j$, $\ttha_i K_{ij} = K_{ij}\ttha_j$, $K^2_{ij} = 1$ and $K_{ij}K_{jk}=K_{jk}K_{ji}=K_{ki}K_{ij}$. The above
models are closely related to the general models in \cite{Minahan:1993} (see also \cite{Brink:1992}) and should hence be integrable. Notice that $K_{ij}$
only acts on the fermionic coordinates whereas  the operators in \cite{Minahan:1993,Brink:1992} also act on the
bosonic coordinates; it is however easy to extend the above models to
this more general setting.  The models (\ref{ccal}) can be
straightforwardly extended to arbitrary $\cN$. The supercharges take
the form
\be
Q_a = \sum_i \tha^i_a\left(p_i + i
\frac{\nu}{2^{\frac{\cN}{2}-1}}\sum_m \prod_{c\neq a}\frac{[(\tha^i_c-\tha^m_c),(\pa_{\tha^i_c}-\pa_{\tha^m_c})]}{x^i{-}x^m}\right)\,,
\ee
where $a,c=1,\ldots,\frac{\cN}{2}$. The hamiltonian has the same form as in (\ref{ccal}), but with $K_{ij}$
given by $K_{ij} =
\frac{1}{2^{\frac{\cN}{2}}}\prod_{c=1}^{\frac{\cN}{2}}[(\tha^i_c -
\tha^j_c),(\pa_{\tha^i_c} - \pa_{\tha^j_c})]$.

The $\cN=4$ models just constructed are conformal, but as we shall see next
the superconformal algebra is not $\mathrm{su}(1,1|2)$. We now turn to the question of what restrictions
follow from demanding $\mathrm{SU}(1,1|2)$ superconformal invariance. With $S=\sum_i\tha_i x_i$ and $\tilde{S} = \sum_i \ttha_i x_i$, the superconformal algebra is satisfied if 
\be
\sum_i x_i W^{[1]}_{ijk} = -\de_{jk}\\, \label{scconstr}
\ee
and $\sum_i x_i \pa_i W^{[0]} = \mathrm{const}$ (the latter condition
follows from $[D,W^{[0]}_i] = -iW^{[0]}_i$, i.e. from conformal
invariance). The other generators are then given by 
\be
J_1 = -\half \sum_i (\tha_i\pa_{\ttha_i} + \ttha_i\pa_{\tha_i})\,,
\quad J_2 = -\mbox{$\frac{i}{2}$}\sum_i (\tha_i\pa_{\ttha_i} -
\ttha_i\pa_{\tha_i})\,, \quad J_3 = \half\sum_i (\ttha_i\pa_{\ttha_i} - \tha_i\pa_{\tha_i})\,,
\ee
and $T= \sum_i x_i\pa_i W^{[0]} - \frac{\cN}{2}$. 
The restriction (\ref{scconstr}) on  $W^{[1]}_{ijk}$, show that the
models (\ref{ccal}) are not $\mathrm{SU}(1,1|2)$ superconformal.  

Another issue is translational invariance. The condition for
superconformal invariance (\ref{scconstr}) is not consistent with
translational invariance of $W^{[1]}$ (since that would imply $\sum_k
\pa_k W^{[1]}=0$). However, after extracting  from $W^{[1]}_{ijk}$ the
non-translational-invariant centre-of-mass part $W^{[1]\rm
cm}_{ijk}=-\frac{1}{MX}$ (where $X=\sum_ix_i$) the
remaining relative part can be taken to be translational-invariant
and the superconformal condition is replaced by $\sum_i x_i
W^{[1]}_{ijk} = -\de_{jk} + \frac{1}{M}$.  With this modification of
the models in (\ref{ccal}), they  
become $\mathrm{SU}(1,1|2)$ superconformal for certain exotic values
of the coupling constant $\nu$, namely when $\nu=\frac{1}{M}$. We will
now address the question of whether there exist $\mathrm{SU}(1,1|2)$
superconformal extensions of the $A_{M-1}$ Calogero models for generic
values of the coupling constant. If we assume that $W^{[0]}$ has an
overall parameter (as in the case of the Calogero models) then (if
demand conformal invariance and discard the above solution) the last equation in (\ref{constr}) decouples into two equations: $\pa_i\pa_j W^{[0]} = \sum_{l}W^{[1]}_{lij}\pa_l W^{[0]}$, and $\pa_i \sum_{n} W^{[1]}_{jnn} = \sum_{l,n}W^{[1]}_{lij}W^{[1]}_{lnn}$. Notice that the latter equation is consistent with (\ref{scconstr}) and conformal invariance.  When $W^{[0]}$ has the Calogero form $W^{[0]} = \frac{\nu}{2} \sum_{n\neq m}\ln|x_n-x_m|$ one can show that there are no solutions to the coupled set of equations (\ref{constr}),(\ref{scconstr}) for $M=3$; we believe that this continues to be true for higher $M$ (it can be shown for all $M>2$ that
for generic $\nu$ there is no solution with two-body interaction
forces only). 
Thus, we conclude that for $M\!>\!2$ there is no natural candidate for an
$\cN=4$ $\mathrm{SU}(1,1|2)$ superconformal $A_{M-1}$ Calogero model which has
the proper $\cN=2$ limit. We would like to stress that this conclusion
depends on the particular (but natural) choice of supercharges (\ref{scharges}).

Is it possible to
find other $\mathrm{SU}(1,1|2)$ superconformal models? 
For simplicity let us discuss the $M=3$ (three-particle) case in more 
detail. There is actually no solution to the set of constraints given
above for any $W^{[0]}$ with an overall parameter and two-body
interactions only, so the conditions have to be
weakened. One has to allow for a $\nu$-independent part in $W^{[0]}$
and/or higher-body interactions if one is to be able to satisfy the
constraints. At least for the $M=3$ case it turns out that it not
sufficient to introduce higher-body interactions, so we will therefore
allow for a $\nu$-independent part in $W^{[0]}$. At this point we recall
that there is another three-particle translational-invariant (bosonic) Calogero model (besides the $A_2$ one) namely the model associated to the $G_2$ Lie algebra \cite{Wolfes:1974}. The hamiltonian is
\be
H_{G_2} = \half\sum_i p_i^2 + \sum_{i<j}\frac{\nu_1(\nu_1{-}1)}{(x_i-x_j)^2}  + 3 \hspace{-4mm}\sum_{\mbox{\scriptsize
$\ba{c} j<k \\ i\neq j \neq k \ea$ }}\hspace{-4mm}\frac{\nu_2(\nu_2{-}1)}{(2x_i-x_j-x_k)^2}\,.
\ee
The two coupling constants $\nu_1$ and $\nu_2$ can be chosen independently. 
This hamiltonian has all the nice properties of the Calogero models, 
such as integrability and exact solvability
\cite{Quesne:1996}. Using reasoning similar to the one
used in the $A_2$ case  one can show that there does not
exist any $\mathrm{SU}(1,1|2)$ superconformal $\cN=4$ extension when
the $Q_a$'s are of the form (\ref{scharges}) and the two coupling constants are unrelated.
Choosing the centre-of-mass part of $W^{[1]}_{ijk}$ as before, allowing for a linear
relation between $\nu_1$ and $\nu_2$, and choosing
\bea
W^{[1]}_{\rm rel} &=& \beta_1\sum_{i<j}(x_i-x_j)^2\ln|x_i-x_j| +
\beta_2 \hspace{-3mm} \sum_{\mbox{\scriptsize $\ba{c} j<k \\ i\neq
j\neq k \ea $}} \hspace{-3mm} (2x_i-x_j-x_k)^2\ln|2x_i - x_j -x_k|\,, \non \\
W^{[0]} &=& \al_1\sum_{i<j}\ln|x_i-x_j| + \al_2 \hspace{-3mm}
\sum_{\mbox{\scriptsize $\ba{c} j<k \\ i\neq j\neq k \ea $}}
\hspace{-3mm} \ln|2x_i - x_j -x_k| \,,
\eea
we have found the following $\mathrm{SU}(1,1|2)$ superconformal
extensions of the $G_2$ model. The following different choices for the parameters
$(\al_1,\al_2,\beta_1,\beta_2)$ are possible: $(-\frac{1}{6} ,\nu_2
,\frac{1}{12},-\frac{1}{12})$,
$(\nu_1,-\frac{1}{6},-\frac{1}{4},\frac{1}{36})$, or $(\frac{1}{3} -
\nu, \nu
,\frac{\nu}{2} - \frac{1}{6},-\frac{\nu}{6})$. The last case is more trivial than the others since it has
$\hat{W}^{[0]}=0$. The corresponding potential is
\bea
V^1_{\rm rel} &=& \sum_{i<j}\frac{1}{x_{ij}^2}\Big[\al_1(\al_1{+}1) -
\al_1(\tha_{ij}\pa_{\tha_{ij}} + \ttha_{ij}\pa_{\ttha_{ij}})  -
2\beta_1 (\tha_{ij}\pa_{\tha_{ij}}\ttha_{ij}\pa_{\ttha_{ij}}\Big]
\non \\ &+& \hspace{-4mm} \sum_{\mbox{\scriptsize $\ba{c} j<k \\ i\neq j\neq k \ea $}} \hspace{-4mm}\frac{1}{x_{ijk}^2}\Big[3\al_2(\al_2{+}1)  - \al_2(\tha_{ijk}\pa_{\tha_{ijk}}
+ \ttha_{ijk}\pa_{\ttha_{ijk}})
-2\beta_2 \tha_{ijk}\pa_{\tha_{ijk}}\ttha_{ijk}\pa_{\ttha_{ijk}}\Big]\,,
\eea
where $x_{ij}= x_i-x_j$, $\tha_{ij}=\tha_i-\tha_j$, and 
$\pa_{\tha_{ij}}= \pa_{\tha_i}-\pa_{\tha_j}$; $x_{ijk}= 2x_i-x_j-x_k$,
$\tha_{ijk}=2\tha_i-\tha_j-\tha_k$, and 
$\pa_{\tha_{ijk}}= 2\pa_{\tha_i}-\pa_{\tha_j}-\pa_{\tha_k}$.

For $M=4$ there also exists a translational-invariant (bosonic) ``Calogero'' model, which in general has two- and four-body interactions \cite{Wolfes:1974b} 
\be
H_{4} = \half\sum_i p_i^2 + \sum_{i<j}\frac{\nu_1(\nu_1{-}1)}{(x_i-x_j)^2}  +\hspace{1mm} 2\hspace{-6mm}\sum_{\mbox{\scriptsize $\ba{c} i< j, k <
l \\ i\neq j\neq k \neq l\ea $}}\hspace{-2mm}\frac{\nu_2(\nu_2{-}1)}{(x_i + x_j-x_k-x_l)^2}\,.
\ee
A similar analysis for this case leads to the following
$\mathrm{SU}(1,1|2)$ superconformal solution with four-body interactions only
\bea
W^{[1]}_{\rm rel} &=& -\frac{1}{8} \hspace{-3mm}\sum_{\mbox{\scriptsize $\ba{c} i< j, k <
l \\ i\neq j\neq k \neq l\ea $}} \hspace{-4mm}(x_i+x_j - x_k
-x_l)^2\ln|x_i+x_j - x_k -x_l|\,, \non \\
W^{[0]} &=& \nu \hspace{-4mm} \sum_{\mbox{\scriptsize $\ba{c} i< j, k <
l \\ i\neq j\neq k \neq l\ea $}} \hspace{-4mm} \ln|x_i+x_j - x_k -x_l| \,, \non \\
V_{\rm rel} &=& \hspace{-6mm} \sum_{\mbox{\scriptsize $\ba{c} i< j, k <
l \\ i\neq j\neq k \neq l\ea $}} \hspace{-3mm}\frac{1}{x_{ijkl}^2}\Big[2\nu(\nu + 1)  -
\nu (\tha_{ijkl}\pa_{\tha_{ijkl}}
+\ttha_{ijkl}\pa_{\ttha_{ijkl}}) +
\frac{1}{4}\tha_{ijkl}\pa_{\tha_{ijkl}}\ttha_{ijkl}\pa_{\ttha_{ijkl}}\Big]
\label{4body}\hspace{0.2mm},
\eea
where, $x_{ijkl}=x_i+x_j-x_k-x_l$, $\tha_{ijkl} =
\tha_i+\tha_j-\tha_k-\tha_l$, and $\pa_{\tha_{ijkl}}=\pa_{\tha_i}+\pa_{\tha_j}-\pa_{\tha_k}-\pa_{\tha_l}$.
One could continue this analysis to higher $M$ and try to find interesting solutions.
One restriction one could impose is that the bosonic part should have
special properties, such as e.g. integrability. Perhaps it is possible
to turn things around and use supersymmetry considerations to construct interesting bosonic models.

\setcounter{equation}{0}
\section{Classical lagrangean treatment} \label{slag}

In this section we perform a study of $\cN=4$ models similar to the one in section \ref{sham}, but from a (classical) lagrangean perspective.

\subsection{One-body models}
In \cite{Ivanov:1989b} (see also \cite{deAzcarraga:1998}) an
$\mathrm{SU}(1,1|2)$ superconformal mechanics model was
constructed. The action is most succinctly written in $\cN=4$
superspace. Our superspace conventions coincide with those of
\cite{Ivanov:1989b} and are as follows:  $D_{a}=\frac{\pa}{\pa
\eta^{a}} + i\bar{\eta}_{a}\pa_t$, $\bar{D}^{a} =
-\frac{\pa}{\pa\bar{\eta}_{a}} -
i\eta^{a}\pa_t$, 
 and $\{D_{a},\bar{D}^{b}\}=-2i\de^{b}_{a}\pa_t$. Indices are
raised and lowered with $\ep_{ab}$ and its inverse $\ep^{ab}$ ($\ep_{ab}\ep^{bc}=\de_{a}^{c}$). To reduce the number of
indices we will sometimes suppress contracted indices with the
understanding that the  first index should be in a ``natural''
position. The action given in \cite{Ivanov:1989b} was constructed in terms of a real superfield with
components $\phi|=x$, $D_{a}\phi| = i\psi_{a}$, $\bar{D}^{a}\phi| =
-i\bar{\psi}^{a}$ and
$[D_{(a},\bar{D}_{b)}]\phi|=F_{ab}$, where $|$ as usual
is shorthand for $|_{\eta_a=0,\bar{\eta}^a=0}$. Since the representation
corresponding to the real superfield $\phi$ is not irreducible one has to constrain the superfield.  The following constraints were used in \cite{Ivanov:1989b}: 
$D^2\phi = -\frac{1}{\phi}D\phi D\phi$, $\bar{D}^2\phi =
-\frac{1}{\phi}\bar{D}\phi\bar{D}\phi$ and $[D_{a},\bar{D}^{a}]\phi
\equiv [D,\bar{D}]\phi = -\frac{2}{\phi}D\phi\bar{D}\phi
+\frac{4\nu}{\phi}$. These constraints are the one-dimensional
analogue of the constraints for the four-dimensional tensor multiplet
\cite{deWit:1982}. The superspace action is
\be
S = \frac{1}{8}\int d t D_{a} D^{a} \bar{D}^{b}\bar{D}_{b}
\left(-\half\phi^2\ln|\phi|\right)\,. \label{1scS}
\ee
After passing to components and eliminating the auxiliary field $F_{ab}$ one obtains the action
\be
S = \half \int d t[ \dot{x}^2 -i\bar{\psi}\dot{\psi} + i\dot{\bar{\psi}}\psi - \frac{(\nu + \bar{\psi}\psi)^2}{x^2} ] \,.
\ee
For completeness we now briefly describe how to pass to the hamiltonian form and
then to quantum mechanics. The classical hamiltonian is $H_{\rm c} =
xp + p_{\psi}\psi + p_{\bar{\psi}}\bar{\psi} - L $, where 
$p=\frac{\de L}{\de x}$, $p_{\psi_{a}} = \frac{\de L}{\de \psi_{a}}$,
and $p_{\bar{\psi}^{a}} = \frac{\de L}{\de \bar{\psi}^{a}}$ are the conjugate momenta (fermionic variational derivatives act from the left). The canonical Poisson brackets are $\{p,x\} = -1$, $\{p_{\psi_{b}},\psi_{a}\}_+=-\de_{a}^{b}$, and $\{p_{\bar{\psi}^{b}},\bar{\psi}^{a}\}_+=-\de^{a}_{b}$. Using  standard methods to deal with the second class constraints $\Ups_a = p_{\bar{\psi}^a} - \frac{i}{2}\psi_a\approx 0$ and $\bar{\Ups}^a = p_{\psi_a} - \frac{i}{2}\bar{\psi}^a\approx 0$, lead to the Dirac brackets $\{\psi_{a},\bar{\psi}^{b}\}^{*} = i\de^{b}_{a}$.  The Noether charges associated to the supersymmetry invariance of the action are $Q_{a}$ and $\bar{Q}^{a}$. At this point we deviate from the particular model discussed so far and assume the supercharges to be of the more general form
\be
Q_{a} = \psi_{a}(p - iw^{[0]}(x)) -
iw^{[1]}(x)\psi_{a}(\bar{\psi}^{b}\psi_{b})\,,\quad 
\bar{Q}_{a} =
\bar{\psi}^{a}(p+iw^{[0]}(x)) +
iw^{[1]}(x)\bar{\psi}^{a}(\bar{\psi}^{b}\psi_{b})\,. \label{1lcharges}
\ee 
In order for $\{Q_{a},\bar{Q}^{b}\}^* = 2i\de^{b}_{a} H_c$ to be
satisfied, the following condition has to be fulfilled: $\pa w^{[0]} =
w^{[0]} w^{[1]}$, and $H_c$ is then determined to be
\be
H_c = \half p^2 + \half (w^{[0]})^2 + \pa w^{[0]} \bar{\psi}^c \psi_c
+ \half \pa w^{[1]}(\bar{\psi}^c \psi_c)^2 \,.
\ee
In the conformal case both $w^{[0]}$ and $w^{[1]}$ are proportional to
$\frac{1}{x}$ and the equation $\pa w^{[0]} = w^{[0]} w^{[1]}$ has two
solutions corresponding to the two solutions found in the quantum case: $w^{[0]}=\frac{\nu}{x}$, $w^{[1]}=-\frac{1}{x}$ and $w^{[0]}=0$, $w^{[1]} = -\frac{2\nu}{x}$, where $\nu$ is a constant.
The next step is to pass to the quantum theory using the usual rule
$\{\cdot,\cdot\}^* \rar i[\cdot,\cdot]$. Since
$[\psi_{a},\bar{\psi}^{b}]_{+} = -\de^{b}_{a}$, the fermions can be
realised as $\psi_{a} = \tha_{a}$ and $\bar{\psi}^{a} =
-\frac{\pa}{\pa\tha_{a}}$. One has to deal with ordering ambiguities
in the supercharges (such ambiguities are absent for $\cN=2$
systems). Requiring that the supercharges still come in
hermitean-conjugate pairs after quantisation (which guarantees that
the hamiltonian is hermitean) and that the supersymmetry algebra is
still satisfied i.e. $[Q_{a},Q^{\dagger b}]_{+} = 2\de^{b}_{a}H$, fixes the ordering ambiguities.  We then regain the supercharges and hamiltonian given earlier in (\ref{1scharges}) and (\ref{1ham}). 

What about superspace formulations for the models corresponding to the
more general supercharges (\ref{1lcharges})? For instance, for the
other conformal model, the superspace action is (when $\nu\neq
\half$)\footnote{When $\nu = \half$ the model is a special case of the
$\mathrm{SU}(1,1|2)$ 
superconformal one and is described by the action (\ref{1scS}) and its
associated constraints with $\nu=0$.}
\be
S = \frac{1}{8}\int d t D_{a} D^{a} \bar{D}^{b}\bar{D}_{b} \frac{1}{1-2\nu}[-\half\phi^2]\,,
\ee
and the constraints are: 
$D^2\phi = -\frac{2\nu}{\phi}D\phi D\phi$, $\bar{D}^2\phi =
-\frac{2\nu}{\phi}\bar{D}\phi\bar{D}\phi$ and $[D,\bar{D}]\phi = -\frac{4\nu}{\phi}D\phi\bar{D}\phi$.
In components the action becomes
\be
S = \half \int dt [\dot{x}^2  - i\bar{\psi}\dot{\psi} + i\dot{\bar{\psi}}\psi - \frac{2\nu}{x^2}(\psi\bar{\psi})^2 ]\,.
\ee
Although the potential has no ``bosonic'' part, the quantum potential has such a part, which, as we have seen, arises from ordering ambiguities. The actions for the models with supercharges (\ref{1lcharges}) can also be written in superspace; the general construction will be given in the next subsection.

\subsection{Many-body models}
The above results will now be extended to many-body systems. In this
section we use the Einstein summation convention: repeated
indices are summed. The construction involves two 
functions, $w^{[0]}_i(x_l)$ and $w^{[1]}_{ijk}(x_l)$, which are assumed to satisfy the following constraints:
\bea
&&w^{[0]}_i = \pa_i w^{[0]}\,, \quad w^{[1]}_{ijk} = \pa_i\pa_j\pa_k
w^{[1]}\,, \non \\
&& w^{[1]}_{li[j}w^{[1]}_{n]ml} = 0 \,, \non \\
&& \pa_i\pa_j w^{[0]} = \pa_l w^{[0]} w^{[1]}_{lij}\,. \label{clconstr}
\eea
The following action
\be
S = \half \int dt [\dot{x}_i \dot{x}^i - i\bar{\psi}_i\dot{\psi}^i +
i\dot{\bar{\psi}}_i\psi^i - (\pa_i w^{[0]})^2 + 2\pa _i\pa_j
w^{[0]}\psi^j\bar{\psi}^k -
\pa_{i}\pa_{j}\pa_{k}\pa_{l}w^{[1]}(\psi^i\bar{\psi}^j)(\psi^k\bar{\psi}^l)]\,,
\label{claction}
\ee
is supersymmetric (see below) if the constraints (\ref{clconstr}) hold.
The associated supersymmetry Noether charges are
\be
Q_a = \psi^i_a(p_i - i \pa_i w^{[0]} - i
\bar{\psi}^{bn}\psi^m_bw^{[1]}_{inm})\,,\quad \bar{Q}^{a} =
\bar{\psi}^a(p_i + i \pa_i w^{[0]} + i
\bar{\psi}^{bn}\psi^m_bw^{[1]}_{inm}) \,,  \label{clcharges}
\ee
and satisfy $\{Q_a, \bar{Q}^{b}\}^*_+ = 2i\de_a^b H_c$, where $H_c$ is
the classical hamiltonian associated to the lagrangean which can be read off from (\ref{claction}).

The conformal $A_{M-1}$ $\cN=4$ Calogero models corresponding to the ones
constructed in the quantum case (cf. (\ref{ccal})) have $w^{[0]}=0$, which means
that classically they have no bosonic potential, however, after passing to
quantum mechanics a bosonic potential is generated as a result of
ordering ambiguities. 

The models (\ref{claction}) can also be written in superspace. To this end we introduce $M$ real superfields $\phi_i$ with components $\phi_i| = x_i$, $D_a\phi_i| = i\psi^i_a$, $\bar{D}^b \phi_i| = -i\bar{\psi}^{b}_i$ and $[D_{(a},\bar{D}_{b)}]\phi_i| = F^i_{ab}$, while the other components have to be constrained.  We introduce the following constraints:
\bea
&&D^2\phi_i = w^{[1]}_{ijk}(\phi_l)D\phi^j D\phi^k\,, \quad \bar{D}^2\phi_i =
w^{[1]}_{ijk}(\phi_l)\bar{D}\phi^j \bar{D}\phi^k \,,\non \\ &&{}[D,\bar{D}]\phi_i
= 2w^{[1]}_{ijk}(\phi_l)D\phi^j\bar{D}\phi^k - 4 w^{[0]}_i(\phi_l)\,. \label{ssconstr}
\eea
In this context the constraints (\ref{clconstr}) can be viewed as
consistency conditions for the superspace constraints
(\ref{ssconstr}). 
We take the superspace action to be of the form
\be
S = \frac{1}{8} \int dt D_aD^a\bar{D}^b\bar{D}_b A(\phi_i)\,,
\ee
where the scalar functional $A$ is assumed to satisfy the equation
\be
\pa_i\pa_j A + w^{[1]}_{ijk} \pa_k A = - \de_{ij} \,.
\ee
The rationale for this choice is that it implies $\bar{D}^2 A = -
\bar{D}\phi_i \bar{D}\phi^i$. The component action can then easily be
obtained  using the
constraints (\ref{clconstr}), (\ref{ssconstr}), with the result
(\ref{claction}). The superspace equation of motion is  
\be
{}[D_{(a},\bar{D}_{b)}]\phi_i = -2 w^{[1]}_{ijk}D_{(a}\phi^j\bar{D}_{b)}\phi^k\,,
\ee
and it can be shown that it reproduces the component equations of motion derived from (\ref{claction}), after the auxiliary fields $F^i_{ab}$ are eliminated. 

The requirement (in addition to  conformal invariance) for $\mathrm{SU}(1,1|2)$ superconformal invariance
is $x^l w^{[1]}_{lij} = -\de_{ij}$. 
The question arises if/how $A$ is related to $w^{[1]}$ and
$w^{[0]}$. One possibility is that $A=w^{[1]}$. This choice is
consistent with the condition for $\mathrm{SU}(1,1|2)$ superconformal
invariance (and in fact implies it if the matrix $M_{ij} = \pa_i\pa_j w^{[1]} +
\de_{ij}$ is invertible and the theory is conformal). Of the many-body models
constructed before in section \ref{sham}, only the classical four-body model corresponding
to (\ref{4body}) satisfies this
constraint. 

From the lagrangean read off from (\ref{claction}) or from the Dirac bracket of the Noether charges (\ref{clcharges}), one can obtain the classical hamiltonian and then pass to quantum mechanics to regain the results obtained in the previous section. The seeming difference between the classical quantities and the quantum ones result from ordering ambiguities. 

\setcounter{equation}{0}
\section{Discussion} \label{sdisc}

Although we only constructed $\cN\geq 4$ extensions of the
$A_{M-1}$ Calogero models, it should also be possible to extend
the results to the Calogero models based on the other root
systems. Another question is whether it is possible to extend the
results to the super-Sutherland
models \cite{Shastry:1993,Brink:1997,Bordner:1999b}.  A more uniform
formulation along the lines of the one in \cite{Bordner:1999b} would also
be desirable. In particular, the integrability properties merit further investigation. The results on exact solvability \cite{Brink:1992,Ruehl:1995,Brink:1997,Gurappa:1999} are
expected to hold also for the supersymmetric extensions (of the models
with discrete spectrum). 

Another issue worth studying is supersymmetry breaking (extending the
results in \cite{Donets:1999}). One could also investigate more
general models, e.g.~by introducing a non-trivial metric so that $H =
\half g^{ij} p_ip_j + V(x_i)$. It may be interesting to try and lift the more general superspace  
constraints (\ref{ssconstr}) for $\phi_i$ to four dimensions, which might lead to a generalisation of the result in \cite{deWit:1982} to many fields.

It was conjectured by Gibbons and Townsend \cite{Gibbons:1998} that an $\cN=4$
$\mathrm{SU}(1,1|2)$ superconformal extension of the $A_{M-1}$ Calogero models could provide a
microscopic description of an extremal $d=4$ Reissner-Nordstr\"om
black hole. This conjecture was partly based on the observation
\cite{Claus:1998,deAzcarraga:1998} that the radial motion of a
super-particle in the near-horizon limit  of a large-mass extremal $d=4$
Reissner-Nordstr\"om black hole is
described by the $\mathrm{SU}(1,1|2)$ superconformal mechanics model
of \cite{Ivanov:1989b}. A related issue is the quantum mechanics of
$M$ slowly moving extremal black holes in four dimensions. This 
multi-black hole mechanics  should be described in terms of $3M$ bosonic and $4M$
fermionic degrees of freedom (just as the multi-black holes in five
dimensions 
discussed in \cite{Michelson:1999b} are described in terms of $4M$ bosonic and $4M$ 
fermionic coordinates). Thus, it would seem that models with $M$ bosonic and $4M$
fermionic coordinates are perhaps more naturally connected with two-dimensional black
holes. However, the near-horizon geometry of an extremal $d=4$ Reissner-Nordstr\"om
black hole is
$\mathrm{adS}_2\times S_2$ and there is a natural ``angular/radial''
split. Hence, it is not excluded that models with $M$ bosonic and $4M$
fermionic coordinates may provide a microscopic description of $d=4$
black holes (this is in the spirit of the
adS/CFT correspondence). Such a many-body model is expected to have an
$\mathrm{SU}(1,1|2)$ superconformal symmetry. For generic
values of the coupling constant, we have not been able to construct an $\cN=4$ extension
of the $A_{M-1}$ Calogero models with an $\mathrm{SU}(1,1|2)$
symmetry. The only possible way around this result is to change the supercharges. Another possibility is that another generalisation of the
one-body case is needed, however, without
further input it is not clear which assumptions should be made to
pinpoint such a model. One criterion one could use \cite{Gibbons:1998}
is that when all
coordinates but one are small, then the model should reduce to the
one-body $\mathrm{SU}(1,1|2)$ model. Even if it turns that
there is no direct connection between the models considered in this paper and
black hole physics, they may still be valuable as toy
models. 

\subsection*{Acknowledgements}

The author would like to thank K. Peeters and P. Saltsidis for
discussions and H. Braden for some useful correspondence. 
This work was supported by the European Commission under contract FMBICT983302.

\appendix
\setcounter{equation}{0}
\section{The $\mathrm{su}(1,1|\frac{\cN}{2})$ algebras}
The $\mathrm{su}(1,1|2)$ Lie superalgebra generators comprise the odd elements $Q_a$, $S_a$ and their hermitean conjugates, together with the even hermitean generators $J_A$, $H$, $D$, and $K$.  A list of supercommutators sufficient to specify the algebra completely is:
\be
\ba{rclcrcl}
{}[Q_a,Q^{\dagger b}]_{+} &=& 2\de_a^bH\,, && [S_{a},S^{\dagger b}]_+&=&2\de_a^bK \non \\
{}[Q_a,Q_b]_+&=&0\,, && [S_a,S_b]_+&=&0 \,,  \non \\
{}[Q_a,S^{\dagger b}]_+ &=& 2i{(\si^A)_{a}}^b J_A - 2 \de_a^b D -  i \de_{a}^{b}T\,, && [J_A,Q_a] &=&-\half {(\si_A)_a}^c Q_c \non \\ 
{}[S_a,Q^{\dagger b}]_+ &=& -2i{(\si^A)_{a}}^b J_A - 2 \de_a^b D +  i \de_{a}^{b}T \,, && [J_A,S_a] &=&-\half {(\si_A)_a}^c S_c \non \\
{}[Q_a,D]&=&\mbox{$\frac{i}{2}$}Q_a\,, && [Q^{\dagger a},D]&=&\mbox{$\frac{i}{2}$}Q^{\dagger a} \non \\ 
{} [S_a,D]&=&-\mbox{$\frac{i}{2}$}S_a\,, && [S^{\dagger a},D]&=&-\mbox{$\frac{i}{2}$}S^{\dagger a}\,, \non \\ {}[K,D] &=& -iK \,, && [H,D]&=&iH \,, \non \\
{}[H,K]&=&2iD \,, && [J_A,J_B]&=&i\ep_{ABC}J_C\,,  \non \\ 
{}[K,Q_a] &=& iS_a \,, && [K,Q^{\dagger a}]&=&iS^{\dagger a} \,, \non \\ 
{}[H,S_a]&=&-iQ_a \,, && [H,S^{\dagger a}]&=& -i Q^{\dagger a}  
\label{su112} \ea 
\ee
Here $a,b=1,2$; $A,B,C=1,2,3$, and the $2\times 2$ matrices ${(\si_A)_a}^b$ are generators of $\mathrm{su}(2)$ and are in this
paper taken to be the usual Pauli matrices satisfying the 
relation $[\si_A,\si_B]=2i\ep_{ABC}\si_C$. In (\ref{su112}) $T$ is a central element which can be removed, however, such a
central extension is present for some of the models considered in
this paper.

The generators of the Lie superalgebra $\mathrm{su}(1,1|\frac{\cN}{2})$ comprise the odd
elements $Q_a$, $S_a$ and their hermitean conjugates, together with the
even generators $J_a{}^b$, $U$, $H$, $D$, and $K$. 
The supercommutators of $\mathrm{su}(1,1|\frac{\cN}{2})$ are the same as in (\ref{su112}) with the following differences. Now the indices
take the values $a,b=1,\ldots, \frac{\cN}{2}$ and the $\si^A$'s are
replaced by $\frac{\cN}{2}\times\frac{\cN}{2}$ matrix generators of $\mathrm{su}(\frac{\cN}{2})$. In this
paper we use the following realisation of these generators: 
$(\la_a{}^b)_c{}^d = -2\de_a^d\de_c^b$ (when $a\!\neq\! b$) and $(\la_a{}^a)_c{}^d =
-2(\de_a^d\de_c^a - \frac{2}{\cN}\de_c^d)$. They satisfy
$[\la_a{}^b,\la_c{}^d] = 2(\de_b^c\la_{a}{}^d -
\de_d^a\la_{c}{}^b)$. Furthermore, the anticommutation relations between the $Q$'s
and the $S$'s are replaced by
\bea
[Q_a,S^{\dagger b}]_+ &=& i(\la_c{}^d)_{a}{}^b J_d{}^c + 2i\de_{a}^bU
- 2 \de_a^b D -  i \de_{a}^{b}T\,, \non \\
{}[S_a,Q^{\dagger b}]_+ &=& -i(\la_c{}^d)_{a}{}^b J_d{}^c - 2i\de_a^bU - 2 \de_a^b D +  i \de_{a}^{b}T \,. 
\eea
The commutation relations involving the additional u(1) generator $U$
are
\be
\ba{rclcrcl}
{}[U,S_a]&=&\half S_a\,, &\phantom{\Big(}& [U,S^{\dagger a}]&=& -\half S^{\dagger
a}\,, \non \\ {}[U,Q_a]&=&\half Q_a\,, && [U,Q^{\dagger a}]&=& -\half
Q^{\dagger a}\,, 
\ea
\ee
and finally, $[J_a{}^b,J_c{}^d] = (\de_b^c J_a{}^d - \de_a^d J_c{}^b)$.

\begingroup\raggedright\endgroup

\end{document}